\documentclass[prd,showpacs,nofootinbib,onecolumn]{revtex4}
\usepackage{graphicx,color}

\begin{document}

\title{Distinguishing two kinds of scalar mesons from heavy meson decays}
\author{ Wei Wang~\footnote{Email: wwang@ihep.ac.cn}$^{a,b}$ and Cai-Dian L\"u~\footnote{Email: lucd@ihep.ac.cn}$^a$
 }

\affiliation{
 $^a$ Institute of High Energy Physics,  Chinese Academy
 of Sciences, Beijing 100049, People's Republic of China \\
 $^b$ Istituto Nazionale di Fisica Nucleare, Sezione di Bari, Bari 70126, Italy}
\date{\today}
\begin{abstract}
In the SU(3) symmetry limit, semileptonic $D^+\to Sl^+\nu$ and
$B^-\to Sl^-\bar\nu$ decays, with $S=a_0(980)$, $f_0(980)$ and
$f_0(600)$, are found to obey different sum rules in the $\bar qq$
and the tetra-quark descriptions  for scalar mesons. Thus these sum
rules can distinguish the two scenarios for light scalar mesons
model-independently. This method also applies to the $\bar B^0\to
J/\psi(\eta_c) S$ decays. Two kinds of SU(3) symmetry breaking
effects are found to be under control, which will not spoil our
method. The branching fractions of the $D^+\to Sl^+\nu$, $B^-\to
Sl^-\bar\nu$ and $\bar B^0\to J/\psi(\eta_c) S$ decays roughly have
the order $10^{-4}$, $10^{-5}$ and $10^{-6}$, respectively. The
ongoing BES-III and the forthcoming Super B experiments are able to
measure these channels and accordingly to provide  detailed
information of the scalar meson inner structure.
\end{abstract}

\maketitle %

\section{Introduction}

Understanding the internal structure of scalar mesons is of prime
interest in hadron physics for several decades. It plays a crucial
role to understand the chiral symmetry breaking mechanisms of the
QCD and the confinement of hadrons. In spite of the striking success
of QCD theory for strong interaction the underlying structure of the
light scalar mesons is still under
controversy~\cite{Spanier:2008zz,Godfrey:1998pd,Close:2002zu,Jaffe:2004ph,Pennington:2007eg}.
The classification  of scalar mesons  suffers from large   hadronic
uncertainties arising from the fact that scalar mesons share the
same spin-parity quantum numbers $J^{PC}=0^{++}$ with the QCD
vacuum. For instance irrespective of the dispute on the existence of
$f_0(600)$ and $K^*_0(800)$ mesons, scalar mesons have been
identified as ordinary $\bar qq$ states, four-quark states or
meson-meson bound states or even those supplemented with a scalar
glueball. In reality, the nonperturbative QCD fluctuations induce
the mixing between different content, which will add further
complexities.

At present, there are already many experimental studies on the
production of scalar mesons in nonleptonic $D$ decays. For instance
branching ratios (BRs) of $D^+\to f_0(600)\pi^+$ and $D^+\to
f_0(980)\pi^+$ have the order of $10^{-3}$ and $10^{-4}$,
respectively~\cite{Amsler:2008zzb}. On the theoretical side
nonleptonic $D$ decays  receive large  hadronic uncertainties, which
can hinder us from getting a clear view of the internal structures
of scalar mesons. On the contrary semileptonic $D^+\to S l^+ \nu$
decays only contain one scalar meson in the final state, which can
be better candidates to probe different structure scenarios of
scalar mesons.

In this work, we propose a model-independent way to distinguish two
different descriptions for scalar mesons, i.e.  the two-quark and
the four-quark scenarios, through the semileptonic $B^-\to
Sl\bar\nu$ and/or $D^+\to Sl^+\nu$ decays, where $S$ denotes a
scalar meson among $a_0(980),f_0(980)$ and $f_0(600)$. In the
following method the flavor SU(3) symmetry will be used to derive
different sum rules for the BRs under these two scenarios. These
semileptonic heavy meson decays are clean as they do not receive
much pollution from the hadronic interactions like nonleptonic heavy
meson decays. In $B$ decays, the lepton pair can also be replaced by
a charmonium state since they own the same properties in the flavor
SU(3) space. For instance, the mode $\bar B^0\to J/\psi S$ is
probably much easier for the experiments to observe.

The layout of this work is given as follows. In
Sec.~\ref{sec:scenarios}, we will briefly discuss different
descriptions for scalar mesons. The SU(3) relations for the
productions rates and the symmetry breaking effects are discussed in
Sec.~\ref{sec:SU(3)} and Sec.~\ref{sec:SU(3)-breaking},
respectively. The last section contains a couple of remarks  and our
conclusion.

\section{Two scenarios for Scalar mesons}
\label{sec:scenarios}

A number of scalar mesons have been experimentally discovered in
different processes, however in contrast to the pseudoscalar  and
vecor mesons the identification of scalar mesons is more difficult
because of their large decay widths causing a strong overlap between
different resonances and background. In particular, in recent years
there has been controversies about the existence of the two light
and very broad states: $K^*_0(800)$ (also refereed as $\kappa$)
meson in the $700- 900$ MeV region; $f_0(600)$ (also denoted as
$\sigma$) meson in the region of $400-1200$ MeV.

The $f_0(600)$ meson, having the same quantum number with the QCD
vacuum,  might play a significant role in the chiral symmetry
breaking and the mass origin of the pseudoscalar mesons. This meson
might also have some relation with the scalar meson proposed in the
linear sigma model at 50 years ago~\cite{Nambu:1960xd}. There have
been many studies on the possible scalar resonance structure  in
different experimental processes.  Although the data of the $2\pi$
invariant mass spectra in the $p\bar p$ annihilation do not show a
distinct resonance structure below 900 MeV~\cite{Amsler:1995bf}, the
existence of the $f_0(600)$ meson is allowed in many processes for
instance the $\pi N$ scattering~\cite{Kaminski:1996da},  the
nonleptonic $D^+\to \pi^+\pi^-\pi^+$ decay
channel~\cite{Aitala:2000xu,Link:2003gb,Bonvicini:2007tc}, and the
$J/\psi\to \omega \pi^+\pi^-$~\cite{Ablikim:2004qna} and
$\psi(2S)\pi^+\pi^-$~\cite{Gallegos:2003gq,Ablikim:2006bz}. This
pole is also derived in the analysis based on chiral symmetry and
Roy equations~\cite{Colangelo:2001df,Caprini:2005zr} and the
analysis using unitarized chiral perturbation
theory~\cite{Caprini:2008fc}.
Refs.~\cite{Pennington:2006dg,Pennington:2007yt} analyzed the
$\sigma\to \gamma\gamma$ process and found that the data are
consistent with a two-step process of $\gamma\gamma\to \pi^+\pi^-$
with a subsequent final state interaction $\pi^+\pi^-\to
\pi^0\pi^0$. This conclusion prevents us from learning anything new
from the coupling of $f_0(600)$ with $2\gamma$. This situation is
also similar for the $K^*_0(800)$ meson. For example,  the data from
the BES-II implies a $\kappa$-like structure in $J/\psi$ decays into
$\bar K^{0*}K^+\pi^-$~\cite{Ablikim:2005ni,Guo:2005wp}.  A number of
phenomenological analysis find a light and broad state consistent
with $K^*_0(800)$ (see many references in the
review~\cite{Spanier:2008zz} in the PDG~\cite{Amsler:2008zzb}) but
this pole is absent in some other
work~\cite{Cherry:2000ut,Kopp:2000gv,Link:2005ge,Aubert:2007dc}. The
inconsistence between different analysis implies the complexities in
the nature of the $f_0(600)$ and $K^*_0(800)$ resonances.

Assuming the existence of the $K^*_0(800)$ and $f_0(600)$, there are
9 mesons together below or near 1 GeV, and in this case it is
reasonable to assume the nonet for scalar mesons below or near 1GeV
consisting of $f_0(600),K^*_0(800),f_0(980)$ and $a_0(980)$. In the
$\bar qq$ picture, scalar mesons are viewed as P-wave
states~\cite{Tornqvist:1995kr}, whose flavor wave functions are
given by
\begin{eqnarray}
 && |f_0(600)\rangle =\frac{1}{\sqrt 2} (|\bar uu\rangle +|\bar
 dd\rangle)\equiv |\bar nn\rangle,\;\;\; |f_0(980)\rangle =|\bar ss\rangle,\\
 && |a_0^0(980)\rangle =\frac{1}{\sqrt 2} (|\bar uu\rangle -|\bar
 dd\rangle),\;\;\; |a_0^-(980)\rangle =|\bar u d \rangle,\;\;\;
 |a_0^+(980)\rangle = |\bar d u \rangle.\nonumber
\end{eqnarray}
In this picture, $f_0(980)$ is mainly made up of $\bar ss$, which is
supported by the large production rates in $J/\psi \to \phi
f_0(980)$ and $\phi\to f_0(980)\gamma$ decays~\cite{Amsler:2008zzb}.
Meanwhile, the experimental data also indicates the nonstrange
component of $f_0(980)$: the BR of $J/\psi \to \omega f_0(980)$ is
comparable with that of $J/\psi\to \phi f_0(980)$.  To accommodate
with the experimental data, $f_0(980)$ is supposed to be the mixture
of $\bar nn$ and $\bar ss$ as
\begin{eqnarray}
 |f_0(980)\rangle &=& |\bar ss\rangle \cos\theta +|\bar nn\rangle
 \sin\theta,\nonumber\\
 |f_0(600)\rangle &=&-|\bar ss\rangle \sin\theta +|\bar nn\rangle
 \cos\theta.
\end{eqnarray}
Using the BRs of $J/\psi \to f_0(980)\phi$ and $J/\psi \to
f_0(980)\omega$, the ratio between the coupling constants of
$f_0(980)K\bar K$ and $f_0(980)\pi\pi$, and the BR  of $\phi \to
f_0(980) \gamma$, the mixing angle $\theta$ is constrained
as~\cite{Cheng:2005nb}
\begin{eqnarray}
 25^\circ<\theta <40^\circ, \;\;\; 140^\circ <\theta
 <165^\circ.\label{eq:mixing-angle-qq}
\end{eqnarray}
Due to the large decay width of $f_0(600)$ meson, the mixing angle
is usually identified as an energy dependent variable. In the above
determination, the mixing angle has been taken as a constant
variable.

From the allowed range of the mixing angle given above, we can see
that $f_0(980)$ is dominated by the $\bar ss$ component. The
expected dominant decay channel $f_0(980)\to K\bar K$  is suppressed
by the small phase space. Then $f_0(980)\to \pi\pi$ decay becomes
dominant arising from some nonperturbative interactions such as the
rescattering mechanism. This is supported by the experimental
data~\cite{Amsler:2008zzb}.

The classical $\bar qq$ picture meets with several difficulties. For
example, since $s$ quark is expected to be heavier than $u/d$ quark,
it is difficult to explain the fact that the strange meson
$K^*_0(800)$ is lighter than the isotriplet mesons $a_0(980)$, and
the isosinglet meson $f_0(980)$ has a degenerate mass with
$a_0(980)$. Moreover, since scalar mesons are identified as the
P-wave states in the $\bar qq$ description, it is difficult to
explain why the $K^*_0(800)$ meson is lighter than its vector
partner $K^*(892)$.

Inspired by these difficulties, other candidate scenarios are
proposed.  In Ref.~\cite{Jaffe:1976ig}, scalar mesons are identified
as diquark-diquark states.  In the SU(3) flavor space, the two
quarks can form two multiplets as
\begin{eqnarray}
 3\otimes 3 =\bar 3 \oplus 6,
\end{eqnarray}
while the other two antiquarks reside in $3$ or $\bar 6$ multiplets.
The diquark in a scalar meson is taken to be totally antisymmetric
for all quantum numbers, color antitriplet, flavor antitriplet, spin
0. The $q^2(\bar q)^2$ states make a flavor nonet, whose  internal
structures  are given as:
\begin{eqnarray}
 |f_0(600)\rangle &=& |\bar uu \bar dd\rangle,\;\;\; |f_0(980)\rangle =|\bar nn\bar ss\rangle,\\\nonumber
 |a_0^0(980)\rangle &=& \frac{1}{\sqrt 2} |(\bar uu-\bar dd)\bar ss\rangle,\;\;\; |a_0^+(980)\rangle =|\bar d u \bar ss\rangle, \;\;\;
 |a_0^-(980)\rangle =|\bar ud \bar ss\rangle. 
\end{eqnarray}
Taking the mixing into account, the isosinglet mesons are expressed
as
\begin{eqnarray}
 |f_0(980)\rangle&=& |\bar nn\bar ss \rangle\cos\phi +|\bar uu \bar
 dd\rangle\sin\phi,\nonumber\\
 |f_0(600)\rangle&=& -|\bar nn\bar ss \rangle\sin\phi +|\bar uu \bar
 dd\rangle \cos\phi,
\end{eqnarray}
where the $\phi$  is constrained as~\cite{Maiani:2004uc}
\begin{eqnarray}
  \phi=(174.6^{+3.4}_{-3.2})^\circ.\label{eq:mixing-angle-qqqq}
\end{eqnarray}

Apart from the $\bar qq$ and the tetraquark picture
(diquark-diquark), there exists another promising candidate
interpretation of scalar mesons: they are molecule states.
Ref.~\cite{Weinstein:1982gc} found within a potential model that
scalar tetraquark systems can only appear in the form of  the $K\bar
K$ molecules . The pair of the isodoublet kaons can form four
different states: three mesons as an isovector and one isosinglet
meson, which can easily explain the degeneracy of the mass of
$a_0(980)$ and $f_0(980)$. More importantly the $K^*_0(800)$ and
$f_0(600)$ mesons are irrelevant and thus this scheme is also
consistent with the absence of these two resonances as physical
states. The molecule description has also been successfully applied
in different processes for instance the effective Hamiltonian
approach within the molecule picture has been used to compute the
$f_0(980)\to\pi\pi$ and $f_0(980)\to \gamma\gamma$
decays~\cite{Branz:2007xp}. It is worthwhile mentioning that our
proposed method in the following section is invalid under this
interesting picture since  the production rates of $f_0(600)$ will
be used.

The identification of the scalar mesons below 1 GeV  as the
four-quark states or molecule states can raise the question about
the scalar $\bar qq$ states. The experimentalists have observed
several scalar mesons above 1GeV. The expectation from the naive
quark model that $\bar qq$ scalar mesons $(L=1)$ are expected to be
heavier than the vector $\bar qq$ partners $(L=0)$ may also give us
a hint that the scalar $\bar qq$ nonet is made of the several
heavier mesons: the isovector and isodoublet mesons $a_0(1450),
K_0^*(1430)$, the isosinglet mesons $f_0(1370)$, $f_0(1500)$,
$f_0(1710)$ with some ambiguities in the choice of the ninth member
since the isosinglet mesons can mix with the scalar gluebll state
through the QCD interactions. Nevertheless, since this work mainly
concerns the scalar mesons below 1 GeV, we will refrain from more
discussions of the nature of this heavier $\bar qq$ nonet.

\section{SU(3) relations }
\label{sec:SU(3)}

\begin{figure}
\begin{center}
\includegraphics[scale=0.4]{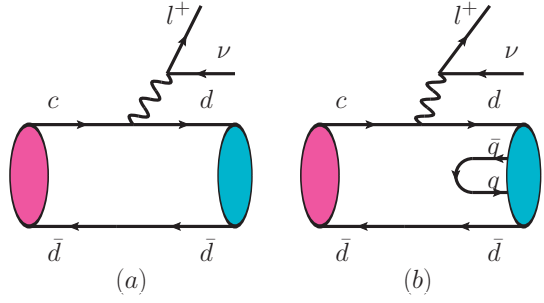}
\end{center}
\caption{Feynman diagrams of $D^+$ decays into a scalar meson. The
left diagram is for two-quark picture, while the right diagram is
for the four-quark mesons. } \label{diagram:Feyn}
\end{figure}

Although the nature of scalar mesons is very complicated, we will
focus on two simple pictures and analyze their productions in the
heavy meson decays. Feynman diagrams for $D^+\to S l^+\nu$ decays in
two different pictures are given in Fig.~\ref{diagram:Feyn}.  The
left panel is for the two-quark scenario, while the right panel is
for the four-quark scenario. If a scalar meson is made of $\bar qq$,
the light quark is generated from the electroweak vertex and the
antiquark  $\bar d$ serves as a spectator. Thus only the component
$\bar d d$ contributes to semileptonic $D$ decays. In the SU(3)
symmetry limit, decay amplitudes of $D\to f_0(980)(f_0(600))l\nu$
channels under the $q\bar q$ picture have the following relation
\begin{eqnarray}
 {\cal A}(D^+\to f_0(980)l^+\nu)&=&-\sin\theta \hat {\cal A},\nonumber\\
 {\cal A}(D^+\to f_0(600) l^+\nu)&=&-\cos\theta\hat {\cal A},
\end{eqnarray}
where the transition amplitude $ \hat {\cal A}$ is defined as
\begin{eqnarray}
 \hat {\cal A}\equiv {\cal A}(D^+\to a_0^0(980)l^+\nu).
\end{eqnarray}
This leads to a sum rule
\begin{eqnarray}
 {\cal B}(D^+\to a_0^0(980)l^+\nu)=
 {\cal B}(D^+\to f_0(980)l^+\nu)+
 {\cal B}(D^+\to f_0(600) l^+\nu),
 \label{sumrule1}
\end{eqnarray}
which is independent of the mixing angles. One may worry about the
accuracy of our above result because of the possible large QCD
scattering effect. However, if we use the hadron picture, we can
still get the same result. The $d\bar d$ pair produced from the weak
interaction in Fig.\ref{diagram:Feyn}(a) can form isospin 0 and
isospin 1 states with the ratio of 1:1. The ratio is directly
obtained from the Clebsch-Gordan coefficients. Although the QCD
scattering can mix between states, the non-perturbative QCD
interactions conserve the isospin. Therefore the sum of production
rates of isospin 0 states on the right hand side of
eq.(\ref{sumrule1}) is always equal to production rates of the
isospin 1 states on the left hand side of eq.(\ref{sumrule1}). The
isospin breaking effect in strong interaction is negligible.

If a scalar meson is composed of four quarks, besides the light
quark from the electroweak vertex and the spectator, another $\bar
qq$ pair is generated from the QCD vacuum. This quark pair could be
$\bar ss$ or $\bar uu$, where both $\bar dd\bar ss$ and $\bar uu\bar
dd$ contribute. The decay amplitudes are given as
\begin{eqnarray}
 {\cal A}(D^+\to f_0(980)l^+\nu)&=&-(\cos\phi+\sqrt 2\sin\phi)\hat {\cal A},\nonumber\\
 {\cal A}(D^+\to f_0(600) l^+\nu)&=&(\sin\phi-\sqrt 2\cos\phi)\hat {\cal
 A},
\end{eqnarray}
which gives
\begin{eqnarray}
 {\cal B}(D^+\to a_0^0(980)l^+\nu)=\frac{1}{3}[{\cal B}(D^+\to f_0(980)l^+\nu)+
 {\cal B}(D^+\to f_0(600) l^+\nu)].\nonumber\\
\end{eqnarray}

It is meaningful to define the ratio of partial decay widths
\begin{eqnarray}
 R=\frac{{\cal B}(D^+\to f_0(980)l^+\nu)+
 {\cal B}(D^+\to f_0(600) l^+\nu)}{{\cal B}(D^+\to
 a_0^0(980)l^+\nu)}.\label{rule1}
\end{eqnarray}
The ratio is 1 for the two-quark description, while it is 3 for the
four-quark description of scalar mesons. Similarly for semileptonic
$B\to Sl\bar\nu$ decays, the charm quark in Fig.\ref{diagram:Feyn}
is replaced by a bottom quark and the $\bar d$ quark is replaced by
a $\bar u$ quark, while leptons are replaced by their charge
conjugates. We have the same sum rules
\begin{eqnarray}
 R&=&\frac{{\cal B}(B^-\to f_0(980)l^-\bar\nu)+
 {\cal B}(B^-\to f_0(600) l^-\bar\nu)}{{\cal B}(B^-\to
 a_0^0(980)l^-\bar\nu)} \nonumber\\
 &=&\left\{ \begin{array}{cc}1 & \mbox{two~ quark}\\
 3 & \mbox{tetra-quark} \end{array} .\right . \label{rule2}
\end{eqnarray}

Let's now examine whether all these channels are experimentally
measurable by estimating the branching ratios for individual
channels. {If the mixing angle is close to $\theta=0^\circ$ or
$\theta=90^\circ$ in the two-quark picture (in four-quark scenario,
the mixing angle is $54.7^\circ$ or $144.7^\circ$), either
$f_0(600)$ or $f_0(980)$ meson has small production rates but the
other one should have  large production rates. Neglecting the highly
suppressed channel, the ratio defined in
eq.(\ref{rule1},\ref{rule2}) can still distinguish the two different
scenarios for scalar mesons.}

If the mixing angle is not close to the values discussed in the
above paragraph, all three $D^+\to Sl^+\nu$ channels would have
similar BRs in magnitude. The BR of the semileptonic $D_s\to
f_0(980)$ decay is measured \cite{Ecklund:2009fi} as
\begin{eqnarray}
&& {\cal B}(D_s\to f_0(980)l\bar\nu)\times {\cal B}(f_0(980)\to
 \pi^+\pi^-)\nonumber\\
 &&=(2.0\pm0.3\pm0.1)\times 10^{-3}.
\end{eqnarray}
Thus as an estimation, generic BRs for the cascade $D^+\to Sl^+\nu$
decays are expected to have the order
\begin{eqnarray}
 f(D)=\frac{V_{cd}^2}{V_{cs}^2}\times 2\times 10^{-3} \simeq 1\times
 10^{-4},
\end{eqnarray}
while the mixing effects can modify the BRs by several times. For
instance if the mixing angles given in
Eq.~(\ref{eq:mixing-angle-qq}) and Eq.~(\ref{eq:mixing-angle-qqqq})
are used, one has the estimates of the BRs
\begin{eqnarray}
 &&{\cal B} (D^+\to f_0(980)l^+ \nu) \simeq \frac{1}{2}f(D)\tan^2\theta
 \simeq (0.04-0.35) \times  10^{-4},\nonumber\\
 &&{\cal B} (D^+\to a_0l^+ \nu) \simeq \frac{1}{2} f(D)\frac{1}{\cos^2\theta} \simeq (0.6-0.8) \times  10^{-4},\nonumber\\
 &&{\cal B} (D^+\to f_0(600) l^+ \nu) \simeq \frac{1}{2} f(D) \simeq (0.4-0.6) \times
 10^{-4}\label{eq:estimation-mixing-angle-two-quark}
\end{eqnarray}
in two-quark description; or
\begin{eqnarray}
 &&{\cal B} (D^+\to f_0(980)l^+ \nu)  \simeq (0.19-0.63) \times  10^{-4},\nonumber\\
 &&{\cal B} (D^+\to a_0l^+ \nu) \simeq (0.5-0.54) \times  10^{-4},\nonumber\\
 &&{\cal B} (D^+\to f_0(600) l^+ \nu) \simeq (0.88-1.4) \times  10^{-4}\label{eq:estimation-mixing-angle-four-quark}
\end{eqnarray}
in the four-quark description, where  the mixing angle given in
Eq.~(\ref{eq:mixing-angle-qqqq}) is used but the uncertainty is
increased as $(175\pm10)^\circ$. The luminosity of BES-III
experiment at BEPC II in Beijing is designed as $3\times 10^{32}
{\rm cm}^{-2} {\rm s}^{-1}$. This experiment, starting running since
summer 2008, will accumulate 30 million $D\bar D$ pair per running
year~\cite{Asner:2008nq}. Even we assume the detect efficiency is
only $20\%$, there will be 600 events per running year if the BR is
used as $1\times 10^{-4}$. It is very likely to observe these decay
channels.

%
%
%

\begin{figure}
\begin{center}
\includegraphics[scale=0.4]{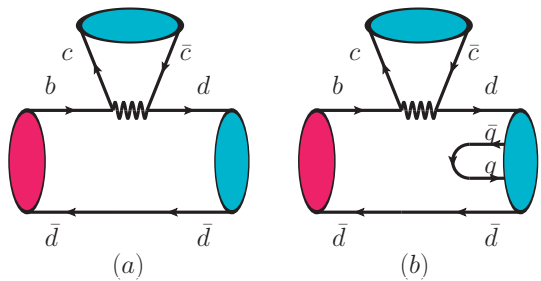}
\end{center}
\caption{Feynman diagrams of $B\to J/\psi(\eta_c) S$ decays. The
left panel is for two-quark picture, while the right panel is for
the four-quark mesons. } \label{diagram:Feyn-2}
\end{figure}

As for the $B$ decays, the generic BR of $B\to Sl\bar\nu$ can be
estimated utilizing the $B\to \rho l\bar\nu$ and $D_s^+\to\phi
l^+\nu$ decays
\begin{eqnarray}
 f_1(B)= {\cal B}(B\to Sl\bar\nu)&\simeq& {\cal B}(B\to \rho l\bar\nu)\frac{{\cal B}(D_s\to f_0(980)l\bar\nu)}{{\cal B}(D_s\to\phi
l\bar\nu)}\nonumber\\
 &\simeq & 10^{-4} \times \frac{10^{-3}}{10^{-2}} =10^{-5},
\end{eqnarray}
where the heavy quark symmetry has been assumed.  The mixing effects
can be similarly analyzed  as in the case of semileptonic $D$
decays. For the three distinct channels, the BRs are  obtained from
Eqs.~(\ref{eq:estimation-mixing-angle-two-quark}) and
~(\ref{eq:estimation-mixing-angle-four-quark}) by changing $f(D)$ by
$f(B)$ and changing $10^{-4}$ by $10^{-5}$. It is also worthwhile
pointing out that the above analysis  is based on the heavy quark
limit $m_{b,c}\to \infty$. The finite heavy quark masses effect
might also provide some changes. Compared with the recently measured
semileptonic $B\to \eta$ decay~\cite{Aubert:2008ct}
\begin{eqnarray}
 {\cal B}(B^-\to \eta l^-\bar\nu)&=&(3.1\pm0.6\pm0.8)\times
 10^{-5},
\end{eqnarray}
we can see that the $B\to S l\bar\nu$ is comparable with the $B\to
\eta l\bar\nu$ decays. Such a large BR offers a great opportunity
for distinguishing the descriptions. Even if the present $B$ factory
does not observe these channels, it is easy for the forthcoming
Super B factory to measure these channels.

The semileptonic $D/B$ decays are clean, which do not receive much
pollution from the strong interaction. But since the neutrino is
identified as missing energy, the efficiency to detect these
channels may be limited.
The lepton pair can also be replaced by some other SU(3) singlet
systems such as a $J/\psi$ or $\eta_c$ meson. Replacing the lepton
pair by the $J/\psi$ and replacing $B^-$ by a $\bar B^0$ state (a
different spectator antiquark will not change the results) in
Eq.~(\ref{rule2}), one can easily obtain the similar sum rules for
the BRs
\begin{eqnarray}
 R&=&\frac{{\cal B}(\bar B^0\to f_0(980)J/\psi)+
 {\cal B}(\bar B^0\to f_0(600) J/\psi)}{{\cal B}(\bar B^0\to a_0^0(980)J/\psi)} \nonumber\\
 &=&\left\{ \begin{array}{cc}1 & \mbox{two~ quark}\\
 3 & \mbox{tetra-quark} \end{array} .\right . \label{rule3}
\end{eqnarray}
%
%
The Feynman diagrams are shown in Fig.~\ref{diagram:Feyn-2}.
Although these channels are hadronic decays, the hadronic
uncertainties are mostly canceled in the sum rule of ratios.  The BR
is expected to have the order
\begin{eqnarray}
 f'(B)= {\cal B}(B\to SJ/\psi)&\simeq& {\cal B}(\bar B^0\to \rho^0 J/\psi)\frac{{\cal B}(D_s\to f_0(980)l\bar\nu)}{{\cal B}(D_s\to\phi
l\bar\nu)}\nonumber\\
 &\simeq & 2.7\times 10^{-5} \times \frac{10^{-3}}{10^{-2}}
 \simeq 3\times 10^{-6}.
\end{eqnarray}
The BRs for the three $B\to SJ/\psi$ decay channels can be obtained
by the results given in
Eqs.~(\ref{eq:estimation-mixing-angle-two-quark}) and
~(\ref{eq:estimation-mixing-angle-four-quark}) with a multiplication
of the factor $0.03$. On experimental side  the $J/\psi$ is easily
detected through a lepton pair $l^+l^-$ and thus these modes may be
more useful. If the $J/\psi$ meson is replaced by $\eta_c$ in
eq.(\ref{rule3}), one can get the similar sum rules.

\begin{figure}
\begin{center}
\includegraphics[scale=0.8]{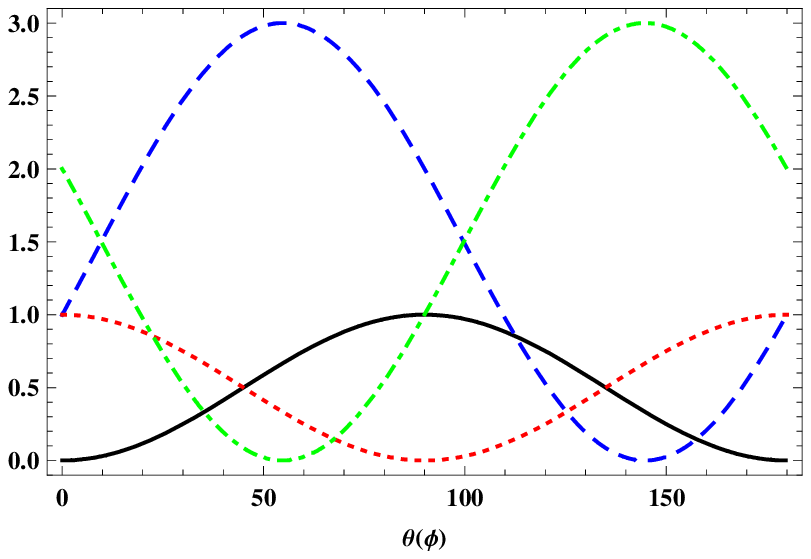}\put (-200,75){R}
\end{center}
\caption{Ratios of branching fractions in $D/B$ decays vs mixing
angles in two-quark and four-quark scenarios. For all three kinds of
decays, the solid (black) and dashed (blue) lines represent ratios $
R_{f_0(980)}$ in the $\bar qq$ and $q^2\bar q^2$ picture,
respectively. The dotted (red) and dot-dashed (green) lines
represent ratios $\kappa R_{f_0(600)}$ in the $\bar qq$ and $q^2\bar
q^2$ picture, respectively.} \label{diagram:mixing-angle}
\end{figure}

With the available data in the future, one can not only distinguish
the different descriptions for scalar mesons but also determine the
mixing angle between the isosinglet mesons. In
Fig.~\ref{diagram:mixing-angle}, we  show  the dependence of ratios
of BRs in $D/B$ decays on the mixing angle in both scenarios.  For
all three kinds of decays, the solid (black) and dashed (blue) lines
represent ratios $ R_{f_0(980)}=\frac{{\cal B}(D/B\to f_0(980) l
\nu)}{{\cal B}(D/B \to a_0^0 l \nu)}$ ($ R_{f_0(980)}=\frac{{\cal
B}(B\to f_0(980) J/\psi)}{{\cal B}(B \to a_0^0 J/\psi)}$ ) in the
$\bar qq$ and $q^2\bar q^2$ picture, respectively. Similarly, we can
define the ratio  $ R_{f_0(600)}=\frac{{\cal B}(D/B\to f_0(600) l
\nu)}{{\cal B}(D/B \to a_0^0 l \nu)}$ ($ R_{f_0(600)}=\frac{{\cal
B}(B\to f_0(600) J/\psi)}{{\cal B}(B \to a_0^0 J/\psi)}$ ).
Unfortunately, due to the large uncertainty of the $f_0(600)$ meson
mass as we will discuss in the following section, we have to modify
the ratio by a phase space factor $\kappa$ for channels involving
$f_0(600)$, especially for the semileptonic $D$ decays. In
Fig.\ref{diagram:mixing-angle}, the dotted (red) and dot-dashed
(green) lines represent ratios $\kappa R_{f_0(600)}$ in the $\bar
qq$ and $q^2\bar q^2$ picture, respectively. The mixing angles in
either scenario are clearly related to the branching ratios.

%


\section{SU(3) symmetry breaking}
\label{sec:SU(3)-breaking}

Remember that the above results are obtained in the flavor SU(3)
symmetry limit. It is also necessary to estimate the symmetry
breaking effects since the strange quark is heavier than $u,d$
quark. In the following we will first analyze the generic size of
symmetry breaking effects and then discuss two different cases.

The large SU(3) symmetry breaking effects in $D$ meson decays, for
example the BR of $D^0\to K^+K^-$ is about three times larger than
$D^0\to \pi^+\pi^-$, may lead to the suspect of the reliability of
the sum rules as in Eq.~(\ref{rule2}). However the hadronic $D^0\to
K^+K^-$ and $D^0\to \pi^+\pi^-$ decays are different with the decay
modes discussed in the this work in several aspects. The former
decays involve two light mesons and the symmetry breaking effects
coming from either of them will be constructive~\footnote{It is
clearer if the factorization method is used although the conclusion
does not depend on this hypothesis. The symmetry breaking is either
from the different $D\to K$ and $D\to \pi$ form factors or decay
constants of kaon and pion. In $D^0\to K^+K^-$ both form factors and
decay constants are larger.}. In semileptonic $B/D$ decay modes the
final state contains only one hadron and the symmetry breaking is
naturally smaller. Moreover in  hadronic $D$ decays, the $D\to
K/\pi$ transitions is induced by the different quark current $c\to
s/d$. On the contrary, the transition current at the quark level in
this work is $c\to d$, same for channels involving isovector and
isosinglet scalar mesons. Accordingly the symmetry breaking effects
in semileptonic $D/B$ decays will be smaller than those in hadronic
$D$ decays.

Generically the size of the SU(3) breaking effect could be roughly
estimated by the mass difference between $u/d$ and $s$ quarks, whose
magnitude is
\begin{eqnarray}
 \frac{m_s-m_{d/u}}{ \Lambda}\sim  0.3
\end{eqnarray}
where $\Lambda$ is the hadronic scale. Taking into this generic
symmetry breaking effect, the two ratios become $R=1\pm0.3$ in the
$\bar qq$ picture and $R=3\pm0.9$ in the tetraquark picture. The
difference between them could be smaller, but they are still
different and our method is still useful. We also expect smaller
SU(3) symmetry breaking effects in $B$ meson decays, since the large
energy release may weaken the effects from the different masses.

One particular origin is that the isospin singlet scalar mesons have
different masses, which can change the phase space in the
semileptonic $D/B$ decays. Fortunately, this SU(3) breaking effect
can be well studied, which almost does not depend on the internal
structure of scalar mesons or the strong interactions. The mass of
$f_0(980)$ is well measured but the mass of $f_0(600)$ meson has
large uncertainties $m_{f_0(600)}=(0.4-1.2)$ GeV. This big range of
masses indeed induces large differences to $D$ decays, since the $D$
meson mass is only 1.87GeV. The BR of the semileptonic decay is
affected by a factor of $(0.31 -5.4)$ depending on the mass of the
$f_0(600)$ meson. Therefore the sum rule in eq.(\ref{rule1}) is not
good unless the $f_0(600)$ meson mass is well measured. But in $B$
meson decays, the sum rule in eq.(\ref{rule2})  will not be affected
sizably, since the $f_0(600)$ meson mass is negligible compared with
the large $B$ meson mass.  Numerically, this correction factor in
$B$ decays is $ (0.9-1.1)$. This will also affect the extraction of
the mixing angle as we have discussed in the previous section.

Another SU(3) breaking effect comes from the decay form factors of
various scalar mesons.  In the two-quark scenario, only the $\bar
dd$ component contributes to the transition form factors shown in
the left diagram of Fig.\ref{diagram:Feyn}.  The  SU(3) symmetry
breaking effect to the form factors is thus negligible. In the
four-quark scenario, the $\bar uu\bar dd$ component in $f_0(980)$
and $f_0(600)$ resonance is different from the internal structure of
$a_0^0$: $\frac{1}{\sqrt 2}(\bar uu-\bar dd)\bar ss$. In the right
diagram of Fig.\ref{diagram:Feyn},  it would be easier to produce
the lighter $\bar uu$ (or $\bar dd$) quark from the vacuum than the
$\bar ss$ quark. The SU(3) symmetry breaking effects may make the
form factor of $D/B \to f_0(600)$ and $D/B \to f_0(980)$ larger than
that of $D/B \to a_0$. It will make the ratio $R$ larger than 3 in
the four-quark scenario. Thus this SU(3) symmetry breaking effects
in the form factors will not spoil our method but instead it will
improve its applicability.

\section{Discussions and conclusion}

In the above two sections, we have discussed the SU(3) relations for
the BRs and the symmetry breaking effects in the heavy meson decays.
The individual production rates for $f_0(600)$ and $f_0(980)$ are
required in our method. The effect in the phase space factor caused
by the large uncertainties in the mass of the $f_0(600)$ can be
directly taken into account. However the physical processes would be
more complicated since both the $f_0(600)$ and $f_0(980)$ need be
reconstructed from the $\pi\pi$ mode~\footnote{$\rho$ is a vector
meson and its decay into $\pi\pi$ occur via the $P$-wave amplitude,
while the decay amplitude of $f_0(600)$ and $f_0(980)$ into $\pi\pi$
belongs to $S$-wave. The background from the $\rho^0\to\pi\pi$ can
be separated within the partial wave analysis. Another possibility
is to use only the $\pi^0 \pi^0$ mode in the analysis, since
$\rho^0$ can not decay to this final state due to the ``wrong" C
parity. Of course, we need more data in this case.}. Considering the
large width of $f_0(600)$, it is difficult to distinguish these two
mesons by the same final state. In this case, the simple ratios
$R_{f_0(980)}$ and $\kappa R_{f_0(600)}$ will be not useful to
constrain the mixing angle. Nevertheless, the ratio $R$ which needs
only the sum of these two states can still provide a method for
distinguishing the two descriptions for scalar mesons.

One more ambiguity comes from the mysterious quark content of scalar
mesons. In this work, we have payed particular attention to the
two-quark and tetra-quark picture. As we have mentioned in the
introduction section the physical situation is more subtle, since
any meson may have two- and four-quark components as part of the
usual Fock state expansion. Those two-quark and four-quark states
are quantum-mechanically mixed, also likely with other potential
candidates with the same quantum numbers. Restricted to the 2-quark
and 4-quark mixtures, two complex decay amplitudes and several
mixing angles are unknown and should  be treated as input
parameters, while only 3 physical observables are available from the
experiments in principle. This mixing problem is not solvable and
the proposed sum rules becomes useless. Nevertheless our method is
at least helpful to rule out one of the possibility. If the ratio
$R$ were close to 1, the pure 4-quark picture is likely to be ruled
out but if the ratio $R$ were close to 3 the pure 2-quark picture is
likely to be excluded.

In conclusion, we have investigated the possibility to distinguish
the two-quark and tetra-quark picture for light scalar mesons. The
semileptonic $D/B\to Sl\bar\nu$ decays and the nonleptonic $B\to
J/\psi(\eta_c) S$ decays are discussed in detail.  These decay
channels have a large potential to be measured on the ongoing
BES-III and the forthcoming Super B experiments. Despite a number of
ambiguities arising from the nonperturbative QCD dynamics or the
SU(3) symmetry breaking effects, our method is useful to distinguish
these two different scenarios.

\section*{Acknowledgement}

W. Wang would like to thank Pietro Colangelo for useful discussions.
This work is partially supported by National Natural Science
Foundation of China under the Grant No. 10735080, 10625525 and
10805037 and Natural Science Foundation of Zhejiang Province of
China, Grant No. Y606252.

\end{document}